\documentclass[bm,aps,
amsfonts,amssymb,
preprint,
nofootinbib]{revtex4}

\usepackage[dvipdfmx]{graphicx}
\usepackage{natbib}
\usepackage{amsmath}

\usepackage{color}

\font\mybb=msbm10 at 12pt

\def\bb#1{\hbox{\mybb#1}}

\def\ZZ {\bb{Z}}

\newcommand\beqa{\begin{eqnarray}}
\newcommand\eeqa{\end{eqnarray}}
\newcommand\n{\nonumber\\}

\begin{document}

{~}

\title{Seiberg-Witten Theory and Monstrous Moonshine
}
\author{
Shun'ya Mizoguchi\footnote[1]{E-mail:mizoguch@post.kek.jp}
}


\affiliation{Theory Center, 
Institute of Particle and Nuclear Studies,
KEK\\Tsukuba, Ibaraki, 305-0801, Japan 
}

\affiliation{SOKENDAI (The Graduate University for Advanced Studies)\\
Tsukuba, Ibaraki, 305-0801, Japan 
}

\begin{abstract} 
We study the relation between the instanton expansion of 
the  Seiberg-Witten prepotential for $D=4$, ${\cal N}=2$ $SU(2)$
SUSY gauge theory for $N_f=0$ and $1$ and the monstrous moonshine.
By utilizing a newly developed simple 
method to obtain the SW prepotential,  it is shown that 
the coefficients of the expansion of  
$q=e^{2\pi \tau}$ in terms of $A^2=\frac{\Lambda^2}{16 a^2}$ ($N_f=0$)
or $\frac{\Lambda^2}{16 \sqrt{2}a^2}$ ($N_f=1$) 
are all integer coefficient polynomials of the moonshine 
coefficients of the modular $j$-function.
A relationship between the AGT 
$c = 25$ Liouville CFT and the  $c = 24$ vertex operator 
algebra CFT of the moonshine module is also suggested.
\end{abstract}

\preprint{KEK-TH-2446}
\date{August 22, 2022}

\maketitle

\newpage
``Seiberg-Witten (SW) theory'' \cite{SW1,SW2} is a method, 
or the entire framework thereof, 
for solving analytically 
(and/or geometrically) the low-energy strongly coupled dynamics of
$D=4$, ${\cal N}=2$ (originally $SU(2)$) SUSY gauge theory, 
based on the idea that, after all, the complexified gauge 
coupling can be identified (for the $SU(2)$ case) 
as the complex structure modulus of 
an elliptic fibration (see \cite{Tachikawalecture} for a review).
It succeeded to determine the exact low-energy effective 
prepotential including the full instanton contributions \cite{NekrasovOkounkov}
in terms of some integrals on the ``Seiberg-Witten curve".
It has developed in connection with various research areas 
of modern string theory and mathematical physics,
such as
embedding into M-theory/Gaiotto duality \cite{MtheoryGaiottoduality},
7-brane system/F-theory \cite{7-braneFtheory},
E-strings \cite{Estrings},
AGT 4d/2d correspondence \cite{AGT} and
matrix models (e.g.)\cite{matrixmodels}.

``Monstrous moonshine'', on the other hand, 
refers to 
the curious fact, first noticed by John McKay in 1979,   
that the Fourier coefficients of the modular $j$-function 
can be written as a simple linear combination of dimensions 
of irreducible representations of the monster group \cite{MonstrousMoonshine}
(see \cite{McKayessay} for a review including 
interesting anecdotes about the monstrous moonshine).
The reasoning for this coincidence was given by constructing $c=24$ vertex operator algebra CFT 
whose character is the $j$-function such that the monster group acts on this module as a symmetry
\cite{FLM}. Later it was shown that this was a $\ZZ_2$ 
asymmetric orbifold \cite{DGH}.

In this paper, we study the relation between the two.
In fact, the fact that the two are related is not in itself surprising. 
This is because a SW curve is a rational elliptic surface over the $u$-plane and
the $j$-function is a fundamental function in the theory of elliptic functions.
However, the details of this specific relationship have not been known until now. 
This paper fills this gap.

We use a newly developed method for easily deriving
the $SU(2)$ SW prepotential to show that 
the coefficients of the instanton expansion of  
the prepotential are related to the monstrous moonshine 
in a way specifically explained in the text. 
In particular, it is shown that, in the $N_f=0$ case,  
the coefficients of the expansion of  
$q=e^{2\pi \tau}$ in terms of $A^2=\frac{\Lambda^2}{16 a^2}$ 
are all integer coefficient polynomials of the moonshine coefficients.
A similar thing holds for $N_f=1$.

The idea of the new method is very simple.
The modular $j$-function has an expansion in terms of $q=e^{2\pi i \tau}$ as
\footnote{All the series expansions performed in this paper have been assisted by Mathematica.}
\beqa
j(\tau)&=&\frac{1}{q}+744+196884 q+21493760 q^2+864299970 q^3+20245856256 q^4+O\left(q^5\right).
\label{modularj}
\eeqa
Thus $q$ is conversely expanded in terms of $\frac1j$ as 
\beqa
q&=&{\textstyle \frac{1}{j}+a_0 \left(\frac{1}{j}\right)^2+\left(a_0^2+a_1\right)
   \left(\frac{1}{j}\right)^3+\left(a_0^3+3 a_1 a_0+a_2\right)
   \left(\frac{1}{j}\right)^4+O\left(\left(\frac{1}{j}\right)^5\right) },
\label{qby1/j}
\eeqa
where $a_0=744$, $a_1=196884$, $a_2=21493760$,... are the coefficients of the 
expansion (\ref{modularj}). Note that the coefficient of the $(\frac1j)^k$ term in (\ref{qby1/j})
is an integer given by a $(k-1)$th order homogeneous polynomial of $a_i$ 
if  the ``degree" of $a_i$ is counted as $i+1$.
Since $q=e^{2\pi i \tau}$, we can obtain $2\pi i$ times the prepotential by taking its  logarithm
and integrating it with respect to  $a$ twice.
On the other hand, suppose that the SW curve is given in the Weierstrass form
\beqa
Y^2&=&X^3+f(u)X+g(u),
\eeqa
then its complex structure modulus $\tau$ is found by inversely solving 
the equation 
\beqa
j(\tau)&=&\frac{12^3\cdot4f(u)^3}{4f(u)^3+27g(u)^2}.
\label{j(tau)}
\eeqa
We expand the rhs of this equation (\ref{j(tau)}) by $u$ around $u=\infty$, 
thereby obtain a $1/j$-expansion of  $1/u$. 
Furthermore, 
since the $1/a$-expansion of  $1/u$ is obtained by the period integral on the SW curve, 
we end up with the $1/a$-expansion of  $q$, 
from which the $1/a$-expansion of  the prepotential ${\cal F}$ is obtained.

As a concrete example, let us consider the pure ($N_f=0$) $SU(2)$ ${\cal N}=2$ SYM theory.
The SW curve is given in the quartic-polynomial representation as \cite{HananyOz}
\beqa
y^2&=&C(x)^2-G(x),~~~C(x)=x^2-u,~~~G(x)=\Lambda^4.
\eeqa
The is equivalent to the Weierstrass form  
\beqa
Y^2&=&X^3+f(u) X + g(u),~~~\n
f(u)&=&-\frac{16}3 u^2 + 4\Lambda^4,~~~\label{fNf=0}
\\
g(u)&=&-\frac{128}{27}u^3+\frac{16}3 \Lambda^4 u
\nonumber
\eeqa
The integral of the holomorphic differential along 
the cycle that collapses as $u\rightarrow\infty$ yields 
$\frac{\partial a}{\partial u}$, which can be obtained by, for instance, 
solving the Picard-Fuchs equation  \cite{KLT}.
The result is 
\beqa
\frac{\partial a}{\partial u}&=&\frac{\sqrt{2}}{4\sqrt{u}}
F\left(
\frac 14, \frac 34,1;\frac{\Lambda^4}{u^2}
\right).
\eeqa
Integrating it with respect to  $u$, we find
\beqa
a&=&-\frac{\sqrt{u}}{\sqrt{2}}\sum_{n=0}^\infty  \frac{(4n-3)!!}{4^{2n}(n!)^2}
\left(
\frac{\Lambda^4}{u^2}
\right)^n,
\eeqa
where we have used the infinite series representation of the hyperelliptic function.

Looking at this, it might appear that, except its prefactor, $a$ has a series expansion in 
$\frac{\Lambda^4}{u^2}$ with rational-number coefficients whose 
denominators are integers containing very many prime factors.
This is not case, however, as all the factors of powers of odd prime integers 
in $(n!)^2$ are contained in $(4n-3)!!$ and hence cancel out, leaving only powers 
of $2$  in the denominator. Moreover, these factors of powers of 2 turn out to be 
absorbed if we take the expansion parameter to be $\frac{\Lambda^4}{8u^2}$ 
instead of $\frac{\Lambda^4}{u^2}$. Therefore, defining
\beqa
A\equiv\frac{\Lambda}{4a},~~~U\equiv\frac{\Lambda^2}{8u}, 
\eeqa
$A^2$ is expanded by $U$  as
\beqa
A^2&=&\frac{U}{\left( 
\sum_{n=0}^\infty  \frac{(4n-3)!!}{(n!)^2} (4U^2)^n
\right)^2}\n
&=&
U+8 U^3+168 U^5+5056 U^7+184040 U^9+7525440 U^{11}+332612800 U^{13}\n
&&+15538219520
   U^{15}+756483502440 U^{17}+38023703291200 U^{19}\n
   &&+1960287432256832
   U^{21}+103165644665826816 U^{23}+O\left(U^{25}\right),
\eeqa
where the coefficients are all positive integers.
Note that the factor of $8$ in the denominator of $U$ is 
the smallest one that can absorb all the factors of powers of $2^{-1}$ 
in the expansion. This $U$ is inversely expanded by $A^2$ as
\beqa
U&=&A^2-8 A^6+24 A^{10}-448 A^{14}-4520
   A^{18}-151872 A^{22}-4095296
   A^{26}\n&&-124070400 A^{30}-3886030632
   A^{34}-126167064640 A^{38}-4206822732736
   A^{42}\n&&-143383813565952
   A^{46}+O\left(A^{50}\right).
\label{UbyA^2}
\eeqa

On the other hand, by plugging (\ref{fNf=0}) into (\ref{j(tau)}), 
we have 
\beqa
\frac 1j&=&\frac{U^4 \left(1-64 U^2\right)}{\left(1-48 U^2\right)^3}\n
&=&U^4+80 U^6+4608 U^8+221184 U^{10}+8847360 U^{12}+254803968 U^{14}-782757789696 U^{18}\n
&&-84537841287168
   U^{20}-6763027302973440 U^{22}-476117122129330176 U^{24}+O (U^{25}).\n
   \label{1/jbyU}
\eeqa
Thus, using this in (\ref{UbyA^2}), we obtain an $A^2$-expansion of  $\frac 1j$.
This expansion can be further used in (\ref{qby1/j}) to finally obtain 
an $A^2$-expansion of  $q$. 
We can see from (\ref{qby1/j}), (\ref{1/jbyU}) and (\ref{UbyA^2}) 
that its expansion coefficients are all integer-coefficient 
polynomials of $a_i$'s.  Though we do not present the explicit expression 
for this expansion of $q$, we instead show the expansion of its logarithm:
\beqa
2\pi i \tau &=& \log A^8+48 A^4
+(a_0+96) A^8
+48 (a_0-304) A^{12}
\n&&
+\frac{1}{2} \left(a_0^2+2496 a_0+2 a_1-1570368\right)
  A^{16}
+\frac{48}{5} \left(5 a_0^2+880 a_0+10 a_1-3352464\right)
  A^{20}
\n&&+\frac{1}{3} \left(a_0^3+7200 a_0^2+6 a_1 a_0-3447648
   a_0+14400 a_1+3 a_2-3648416256\right) A^{24}
\n&&+\frac{48}{7} \left(7
   a_0^3+9968 a_0^2+42 a_1 a_0-12245520 a_0+19936 a_1+21 a_2-6513833472\right)
   A^{28}
\n&&+\frac{1}{4} \left(a_0^4+14208 a_0^3+12 a_1 a_0^2+140160
   a_0^2+85248 a_1 a_0+12 a_2 a_0\right.
\n&& 
\left.
-15707695104 a_0+6 a_1^2+280320 a_1+42624 a_2+4
   a_3-6519374734464\right) A^{32}+O\left(A^{34}\right).\n
\label{2piitau}
\eeqa

Thus we can find the expansion of $2\pi i$ times the prepotential ${\cal F}$ by   
integrating (\ref{2piitau}) with respect to  $a$ twice. For example, if we integrate the second term 
$48 A^4$ with respect to $a$ twice, we get
\beqa
48 A^4
&\stackrel{(\int da)^2}\rightarrow&
\frac{48}{4^4\cdot 3\cdot2}\frac{\Lambda^4}{a^2}=\frac1{32}\frac{\Lambda^4}{a^2},
\eeqa
which agrees with the $k=1$ term of the known $N_f=0$ prepotential 
\beqa
{\cal F}={i\, a^2\over 2 \pi} \left( 
4\log {a\over \Lambda}
- 6 +
8 \log 2 -
\sum_{k=1}^\infty {\cal F}_k \left(\Lambda\over a\right)^{4
k}\right)
\label{F_Nf=0}
\eeqa
(Table \ref{Table1}).
From the third term $(a_0+96) A^8$ we can compute the $k=2$ term. 
By using $a_0=744$ we find 
\beqa
(a_0+96) A^8
&\stackrel{(\int da)^2}\rightarrow&
\frac{a_0+96}{4^8\cdot 7\cdot6}=\frac5{16384}
\frac{\Lambda^8}{a^6}=\frac5{2^{14}}\frac{\Lambda^8}{a^6},
\eeqa
which is also the correct result.

In fact, $a_0=744$ ($=3\times\mbox{dim}E_8$) has nothing to do 
with the monster; the first Fourier coefficient $a_1$ related 
to monster representations appears for the first time in the $A^{16}$ term.
From this, by using $a_0=744$, $a_1=196884$ and 
integrating with respect to  $a$ twice, we find
\beqa
\frac12(-1570368+2496 a_0+a_0^2+2 a_1) A^{16}
&\stackrel{(\int da)^2}\rightarrow&
\frac{\frac12(-1570368+2496 a_0+a_0^2+2 a_1)}{4^{16}\cdot 15\cdot14}
\frac{\Lambda^{16}}{a^{14}}
\n&=&\frac{1469}{2^{31}}
\frac{\Lambda^{16}}{a^{14}},
\eeqa
which is a correct answer. In this way, all the known results can be 
correctly recovered. 
Incidentally, since this method directly determines $\tau$, 
it also correctly produces the perturbative part of ${\cal F}$ (\ref{F_Nf=0}).
Indeed, $\frac 1{2\pi i}\times$ 
\beqa
\int da\int da \log A^8 &=&6 a^2+\frac{1}{2} a^2 \log \left(\frac{\Lambda ^8}{2^{16} a^8}\right)
\eeqa
coincides with the perturbative part.

\begin{table}[tbp]
\centering
\begin{tabular}{|ccc|}
\hline
~$k$~&
Rational-coefficient polynomial of $a_i$'s
&${\cal F}_k$\\
\hline 
1& $\frac{48}{4^4\cdot 3\cdot2}$ &$1\over 2^5$\\
2& $\frac{a_0+96}{4^8\cdot 7\cdot6}$ &$5\over 2^{14}$\\
3& $\frac{48(a_0-304)}{4^{12}\cdot 11 \cdot 10}$ &$3\over 2^{18}$\\
4& $\frac{\frac{1}{2} \left(a_0^2+2496 a_0+2 a_1-1570368\right)}{4^{16}\cdot 15\cdot 14}$ &$1469\over 2^{31}$\\
5&
$\frac{\frac{48}{5} \left(5 a_0^2+880 a_0+10 a_1-3352464\right)}{4^{20}\cdot 19\cdot 18}$
&$4471\over 2^{34} \cdot 5$ \\
6&
$\frac{\frac{1}{3} \left(a_0^3+7200 a_0^2+6 a_1 a_0-3447648 a_0+14400 a_1+3
   a_2-3648416256\right)}{4^{24}\cdot 23\cdot 22}$
&$40397\over 2^{43}$  \\
7&
$\frac{\frac{48}{7} \left(7 a_0^3+9968 a_0^2+42 a_1 a_0-12245520 a_0+19936 a_1+21
   a_2-6513833472\right)}{4^{28}\cdot 27\cdot 26}$
&$441325\over  2^{47}\cdot 7$   \\
8&
$\frac{\frac{1}{4} \left(a_0^4+14208 a_0^3+12 a_1 a_0^2+140160 a_0^2+85248 a_1 a_0+12
   a_2 a_0-15707695104 a_0+6 a_1^2+280320 a_1+42624 a_2+4 a_3-6519374734464\right)}{4^{32}\cdot 31\cdot 30}$
&$866589165\over 2^{64}$\\
$\vdots$&$\vdots$&$\vdots$\\
\hline
\end{tabular}
\caption{\label{Table1} The instanton expansion of the prepotential for $N_f=0$ 
and the monstrous moonshine.
If we use the actual values of the Fourier coefficients $a_i$'s of the $j$-function
in the middle column, we re-derive the correct ${\cal F}_k$ in the right 
column on the corresponding row.  
Since each $a_i$ is a integer-coefficient linear combination of the dimensions 
of irreducible representations of the monster, ${\cal F}_k$ is also a 
rational-coefficient polynomial of them.}
\end{table}

The similar is true for $N_f=1$.
The SW curve for $N_f=1$ is 
\beqa
y^2&=&C(x)^2-G(x),~~~C(x)=x^2-u,~~~G(x)=\Lambda^3(x+m)
\eeqa
in the quartic-polynomial representation, whose equivalent 
Weierstrass form reads 
\beqa
Y^2&=&X^3+f(u,m) X + g(u,m),~~~\n
f(u,m)&=&-\frac{16}3 u^2 + 4\Lambda^3 m,~~~\label{fNf=1}
\\
g(u,m)&=&-\frac{128}{27}u^3+\frac{16}3 \Lambda^3 m u -\Lambda^6.
\nonumber
\eeqa
From these data, $1/j$ can be computed as  
\beqa
\frac1j&=&
\frac{U^3 \left(\left(64 \hat{m}^3+432\right) U^3-72 \hat{m} U^2-\hat{m}^2
   U+1\right)}{\left(48 \hat{m} U^2-1\right)^3}\n
&=&-U^3+\hat{m}^2 U^4-72 \hat{m} U^5+\left(80 \hat{m}^3-432\right) U^6-3456 \hat{m}^2
   U^7\n
   &&+\left(13824 \hat{m}^4-144 \hat{m} \left(64 \hat{m}^3+432\right)\right) U^8-110592
   \hat{m}^3 U^9\n&&
   +\left(1105920 \hat{m}^5-13824 \hat{m}^2 \left(64
   \hat{m}^3+432\right)\right) U^{10}\n&&
   +\left(79626240 \hat{m}^6-1105920 \hat{m}^3 \left(64
   \hat{m}^3+432\right)\right) U^{12}+382205952 \hat{m}^5 U^{13}
   \n&&
   +\left(5350883328
   \hat{m}^7-79626240 \hat{m}^4 \left(64 \hat{m}^3+432\right)\right) U^{14}+42807066624
   \hat{m}^6 U^{15}\n&&
   +\left(342456532992 \hat{m}^8-5350883328 \hat{m}^5 \left(64
   \hat{m}^3+432\right)\right) U^{16}+O\left(U^{17}\right),
   \label{1/jNf=1}
\eeqa
where $U\equiv\frac{\Lambda ^2}{16 u}$,  $\hat{m}\equiv \frac{4 m}{\Lambda }$ 
in this $N_f=1$ case.

On the other hand, $\frac{\partial a}{\partial u}$ is given by 
using the quadratic and cubic transformations of the hypergeometric functions  \cite{MasudaSuzuki}
\beqa
\frac{\partial a}{\partial u}&=&
\frac{F\left(\frac1{12},\frac5{12},1;\frac{12^3}j
\right)}{\sqrt{2}(-3f(u,m))^{\frac14}}.
\eeqa
Plugging  (\ref{fNf=1}), (\ref{1/jNf=1}) in this equation, we find  
\beqa
\frac{\partial a}{\partial u}
&=&
\frac{1}{2 \sqrt{2u}}(1+12 \hat{m} U^2-60 U^3+420 \hat{m}^2 U^4-5040 \hat{m} U^5+\left(18480
   \hat{m}^3+13860\right) U^6
\n&&-360360 \hat{m}^2 U^7+\left(900900 \hat{m}^4+2162160
   \hat{m}\right) U^8+\left(-24504480 \hat{m}^3-4084080\right) U^9
\n&&
   +\left(46558512
   \hat{m}^5+232792560 \hat{m}^2\right) U^{10}+\left(-1629547920 \hat{m}^4-931170240
   \hat{m}\right) U^{11}
\n&&
+\left(2498640144 \hat{m}^6+21416915520
   \hat{m}^3+1338557220\right) U^{12}
\n&&
+\left(-107084577600 \hat{m}^5-133855722000
   \hat{m}^2\right) U^{13}
\n&&
   +\left(137680171200 \hat{m}^7+1807052247000
   \hat{m}^4+401567166000 \hat{m}\right) U^{14}+O\left(U^{15}\right)).
\eeqa
Integrating this with respect to $u$, we derive a $u$-expansion of $a$, 
from which the expansion of $U\equiv\frac{\Lambda ^2}{16 u}$ 
in terms of $A\equiv\frac\Lambda{4\sqrt{2} a}$ is found as follows:
\beqa
U&=&A^2-8 A^6 \hat{m}+24 A^8+24 A^{10} \hat{m}^2+64 A^{12} \hat{m}-8 A^{14} \left(56
   \hat{m}^3+81\right)+6960 A^{16} \hat{m}^2
   \\&&
   -40 A^{18} \left(\hat{m} \left(113
   \hat{m}^3+1128\right)\right)+384 A^{20} \left(449 \hat{m}^3+268\right)-1344 A^{22}
   \left(\hat{m}^2 \left(113 \hat{m}^3+1517\right)\right)
   \n&&
   +128 A^{24} \hat{m} \left(46735
   \hat{m}^3+77976\right)-8 A^{26} \left(511912 \hat{m}^6+11530560
   \hat{m}^3+2197485\right)+O\left(A^{27}\right).\nonumber
\label{UbyA^2Nf=1}
   \eeqa
Thus $\frac1j$ is expanded by $A^2$, whose coefficients are this time
integer-coefficient polynomials of $\hat m$.
This yields
\beqa
q&=&
-A^6+A^8 \hat{m}^2-48 A^{10} \hat{m}+A^{12}
   \left(a_0+48 \hat{m}^3-504\right)-2 A^{14} \left((a_0+372)
   \hat{m}^2\right)
\n&&
   +A^{16} \left((a_0+1248) \hat{m}^4+96 (a_0-512)
   \hat{m}\right)
\n&&
+A^{18} \left(-192 (a_0-212) \hat{m}^3-a_0^2+1008
   a_0-a_1-61992\right)
\n&&
+3 A^{20} \hat{m}^2 \left(32 (a_0+88)
   \hat{m}^3+a_0^2+928 a_0+a_1-939920\right)
\n&&
-3 A^{22} \left(\hat{m}
   \left(\left(a_0^2+2864 a_0+a_1-1343656\right) \hat{m}^3+48 a_0^2-48896 a_0+48
   (a_1+67320)\right)\right)
\n&&
+A^{24} \left(\left(a_0^2+4800
   a_0+a_1-1149216\right) \hat{m}^6+48 \left(9 a_0^2-3264 a_0+9 a_1-2473280\right)
   \hat{m}^3
\right.\n&&\left.+a_0^3-1512 a_0^2-1512 a_1+3 a_0 (a_1+126000)+a_2-2282112\right)
\n&&
-4
   A^{26} \left(\hat{m}^2 \left(12 \left(9 a_0^2+2640 a_0+9
   a_1-4424600\right) \hat{m}^3+a_0^3+1530 a_0^2
\right.\right.\n&&\left.\left.
 ~~~~~~~~~~+3 (a_1-915340) a_0+1530
   a_1+a_2+215271600\right)\right)
\n&&
+6 A^{28} \left(8 \left(3 a_0^2+2848
   a_0+3 (a_1-583120)\right) \hat{m}^7+\left(a_0^3+4572 a_0^2+3 (a_1-1279560)
   a_0
\right.\right.\n&&\left.\left.
 ~~~~~~~~~~+4572 a_1+a_2-632801280\right) \hat{m}^4+32 \left(a_0^3-1524 a_0^2+3
   (a_1+130008) a_0
\right.\right.\n&&\left.\left.
 ~~~~~~~~~~-1524 a_1+a_2-2995968\right) \hat{m}\right)
\n&&
+A^{30}
   \left(-4 \left(a_0^3+7614 a_0^2+3 (a_1-966660) a_0+7614
   a_1+a_2-2146774992\right) \hat{m}^6
\right.\n&&\left.
 ~~~~~~~~~~ -768 \left(a_0^3-498 a_0^2+(3 a_1-670444)
   a_0-498 a_1+a_2+73856560\right) \hat{m}^3
\right.\n&&\left.
 ~~~~~~~~~~-a_0^4+2016 a_0^3-2 a_1^2-948024 a_1-6
   a_0^2 (a_1+158004)
\right.\n&&\left.
 ~~~~~~~~~~+a_0 (6048 a_1-4 a_2+67052160)+2016
   a_2-a_3-146853000\right)
\n&&+O\left(A^{32}\right),
\eeqa
whose coefficients are again integer-coefficient polynomials of 
$a_i$'s and $\hat m$.
The $a_1$, $a_2$ and $a_3$ related to the monster group 
representation appear for the first time in the coefficients of 
$A^{18}$, $A^{26}$ and $A^{30}$, respectively.
Taking the logarithm and integrating with respect to $a$ twice, 
we can derive $2\pi i$ times the $N_f=1$ prepotential 
\beqa
{\cal F}^{N_f=1}&=&{i\, a^2\over 2 \pi} \left( 
3\log \frac{a}{\Lambda}
+\frac{1}{2} \left(
-9
+15\log 2
+\pi i
\right)
-\frac{m^2 \log \frac a\Lambda}{2 a^2}
-\sum_{k=2}^\infty {\cal F}^{N_f=1}_k \left(\Lambda\over a\right)^{2
k}
\right)\label{F_Nf=1}
\eeqa
with ${\cal F}^{N_f=1}_k$'s shown in Table \ref{Table2}.
\begin{table}[tbp]
\centering
\begin{tabular}{|ccc|}
\hline
~$k$~&
Rational-coefficient polynomial of $a_i$'s
&${\cal F}^{N_f=1}_k$\\
\hline 
2&
$\frac{\frac12(96 {\hat m} - {\hat m}^4)}{(4\sqrt{2})^4\cdot3\cdot 2}$
&$\frac{\tilde  m}{32}-\frac{\tilde m^4}{48}$
\\
3&
$\frac{\frac13(1512 - 3 a_0 - {\hat m}^6)}{(4\sqrt{2})^6\cdot5\cdot 4}$
&$-\frac{3}{8192}-\frac{\tilde m^6}{480}$
\\
4&
$\frac{\frac14( (4 a_0+384)\hat m^2 - {\hat m}^8)}{(4\sqrt{2})^8\cdot7\cdot 6}$
&$\frac{5\tilde m^2}{16384}-\frac{\tilde m^8}{2688}$
\\
5&
$\frac{\frac15( (-240 a_0+124800)\hat m - {\hat m}^{10})}{(4\sqrt{2})^{10}\cdot9\cdot 8}$
&$-\frac{7\tilde m}{393216}-\frac{\tilde m^{10}}{11520}$
\\
6&
$\frac{\frac16(-390096 - 3024 a_0 + 3 a_0^2 + 6 a_1 +(- 87552 
 + 288 a_0)
\hat m^3 - 
\hat m^{12})}{(4\sqrt{2})^{12}\cdot11\cdot 10}$
&$\frac{153}{536870912}+\frac{3\tilde m^3}{262144} -\frac{\tilde m^{12}}{42240}$
\\
7&
$\frac{\frac17(
(7382760 
 - 5208 a_0
 - 7 a_0^2 
 - 14 a_1 
)\hat m^2 - 
\hat m^{14}
)}{(4\sqrt{2})^{14}\cdot13\cdot 12}$
&$-\frac{715 \tilde m^2}{536870912}-\frac{\tilde m^{14}}{139776}
$
\\
8&
$\frac{\frac18(
(-46891008 
- 393216 a_0 
+ 384 a_0^2 
+ 768 a_1) 
\hat m +(- 6281472 
+ 9984 a_0 
 + 4 a_0^2 
 + 8 a_1) 
\hat m^4 - 
\hat m^{16}
)}{(4\sqrt{2})^{16}\cdot15\cdot 14}$
&$\frac{1131\tilde m}{21474836480}
+\frac{1469\tilde m^4}{2147483648} 
-\frac{\tilde m^{16}}{430080}$
\\
9&
$\frac{
{\scriptsize \begin{array}{c}
\frac{1}{9} \left(-864 a_0^2 \hat{m}^3+366336 a_0 \hat{m}^3-1728 a_1
   \hat{m}^3-3 a_0^3+4536 a_0^2-557928 a_0
   \right.\\[-5pt]
   \left.
   -18 a_0 a_1+9072 a_1-9
   a_2-\hat{m}^{18}+381411072 \hat{m}^3+123415488\right)
\end{array} 
}
}{(4\sqrt{2})^{18}\cdot17\cdot 16}$
&
$
-\frac{385}{549755813888}-\frac{525\tilde m^3}{4294967296}-\frac{\tilde m^{18}}{1253376}
$
\\
$\vdots$&$\vdots$&$\vdots$\\
\hline
\end{tabular}
\caption{\label{Table2}  The instanton expansion of the prepotential for $N_f=1$
and the monstrous moonshine. We have defined 
$\tilde m\equiv\frac m\Lambda$ and
$\hat m\equiv\frac {4m}\Lambda$. Note that ${\cal F}^{N_f=1}_k$ in this Table 
is related to ${\cal F}^1_k$ in Appendix D of \cite{Ohta} as 
${\cal F}^1_k
=-\frac{1}{2}\Lambda^{2k}{\cal F}^{N_f=1}_k$(here), $\Lambda_1=\Lambda$(here).
}
\end{table}

In this paper, we have shown that, in both $N_f=0$ and $1$ cases, 
$q$ is expanded by $A^2$ whose coefficients are integer-coefficient 
polynomials of $a_i$'s (as well as $\hat m$ for $N_f=1$) if $A\propto a^{-1}$ 
is appropriately defined. As a result, 
the coefficients of 
the instanton expansion of the prepotential are expressed as rational-coefficient 
polynomials of the ``moonshine coefficients''  $a_i$'s  (as well as $\hat m$ for $N_f=1$)
as shown in Tables \ref{Table1} and \ref{Table2}.
We expect a similar relationship to hold for $N_f=2$ and $3$,
and perhaps even for E-string theory.  

From these tables, we can see that in both cases the polynomials contain 
``inhomogeneous terms", that is, the terms 
that remain after all $a_i$'s are set to $0$.
Of course, these numbers are not arbitrarily determined, 
but are determined by the given SW curves.
In fact, setting all $a_i$'s to $0$ in (\ref{modularj}) amounts to 
approximating $j(\tau)$ by $\frac1q$.
Since
\beqa
j(\tau)=\left(
\frac{\vartheta_3^8+\vartheta_4^8+\vartheta_2^8}{2\eta^8}
\right)^3,
\eeqa
this implies that the whole affine $E_8$ character is replaced solely by 
the contribution of the ground state.
Thus we may conclude that these inhomogenous terms represent 
information on the ground state of the moonshine module. 

We note that something similar occurs with the asymptotic behavior 
of the Atiyah-Hitchin (AH) metric, where the similar replacements 
$\vartheta_3\rightarrow 1$, $\vartheta_4\rightarrow 1$, $\vartheta_2\rightarrow 0$ 
and $\eta\rightarrow q^{\frac1{24}}$ lead to the Taub-NUT metric 
with a negative NUT charge \cite{AH}. The AH space is known to be 
the moduli space of $D=3$ ${\cal N}=2$ $SU(2)$ SUSY gauge theory \cite{SWD=3},
and the instanton corrections resolves the singularity of the negative-charge 
Taub-NUT. Since the Taub-NUT space is the transverse space of the M-theory 
lift of a D6-brane and the AH-space corresponds to an O6-plane, 
our results in the present analysis may be considered to show 
the corresponding facts in the D7/O7 system.

The results in Tables \ref{Table1} and \ref{Table2} 
also show that the coefficients of the polynomials of $a_i$'s 
contain large integers (unlike the moonshine).
Since each term of the instanton expansion of the prepotential 
is known to be expressed as a $c=25$ Liouville correlation function \cite{AGT}, 
while the moonshine module is a $\ZZ_2$ orbifold of 
$c=24$ vertex operator algebra CFT \cite{DGH}, 
we suspect that these mysterious large numbers may be interpreted as a contribution 
coming from the missing $c=1$ CFT.

The author thanks Kazunobu Maruyoshi, Sota Nakajima, Kazuhiro Sakai and 
Taro Tani for valuable discussions.

\end{document}